\documentstyle[11pt,paspconf,epsf]{article}

\begin{document}

\title{SPH P$^3$MG Simulations of the Lyman-$\alpha$
Forest}

\author{J. W. Wadsley and J. Richard Bond}
\affil{Canadian Institute for Theoretical Astrophysics and 
Department of Astronomy, University of Toronto, 
McLennan Physical Laboratories, Toronto,
ONT M5S 1A1, Canada}




\begin{abstract}
Our understanding of the Lyman-$\alpha$ forest has received a great
boost with the advent of the Keck Telescope and large 3D
hydrodynamical simulations. We present new simulations using the SPH
technique with a P$^3$MG (Particle-Particle Particle-MultiGrid)
non-periodic gravity solver.  Our method employs a high resolution (1
kpc) inner volume, essential for capturing the complex gas physics, a
larger low resolution volume, essential for correct larger scale tidal
fields and a self-consistently applied, uniform tidal field to model
the influence of ultra long waves. We include a photoionizing UV flux
and relevant atomic cooling processes. We use constrained field
realisations to probe a selection of environments and construct a
statistical sample representative of the wider universe. We generate
artificial Lyman-$\alpha$ spectra and fit Voigt profiles. We examine
the importance of (1) the photoionizing flux level and history, (2)
tidal environment and (3) differing cosmologies, including CDM and
CDM+{$\Lambda$}. With an appropriate choice for the UV flux, we find
that the data is fit quite well if the rms density contrast is $\sim$ 1
at $z\sim 3$ on galaxy scales.
\end{abstract}


\keywords{lyman alpha clouds,n-body simulations,hydrodynamics}


\noindent
\small{{\sl To appear in} {\it Computational
Astrophysics}, {\sl Proceedings of the 12th Kingston Conference, Halifax,
October 1996, ed. D. Clarke \& M. West (PASP)}}

\section{Introduction}
We hope to be able to use the current wealth of Lyman-$\alpha$
absorption data to constrain the shape and normalisation of the
density power spectrum and thus the cosmological model. The extraction
of this information is complicated by gas physics, the ultraviolet
flux, the star formation history and supernova energy injection. By
employing gasdynamical simulations, we can incorporate the detailed
high-redshift environment of the clouds self-consistently. We perform
our Lyman-$\alpha$ simulations using the SPH method in conjunction
with a gravity solver based on the multigrid method. We describe our
code, with attention to the novel features, in section~\ref{code}
 
Simulations have to be carefully constructed to achieve both the
necessary resolution and simultaneously provide a good sample of the
universe. Large scale power is also of particular importance for
producing the correct tidal environment for Lyman-$\alpha$
clouds. Recent Lyman-$\alpha$ simulations ({\it e.g.} \cite{R96},
\cite{ME96}, \cite{M96}, \cite{dave96} and \cite{Z96}) have traded
good $k$-space resolution $210-17.95$ h Mpc$^{-1}$ kpc against size,
$3-22$ Mpc. We address these design issues in section~\ref{IC} and see
also Bond \& Wadsley (1996), (hereafter BW).

In generating the line statistics of the artificial Lyman-$\alpha$
spectra, it is important to fit the lines in a manner directly
comparable with observations (\cite{ME96}, \cite{dave96}). Our method
of fitting Voigt profiles is described with our results in
section~\ref{results}

\section{Numerical Method}\label{code}

Smoothed Particle Hydrodynamics (SPH) is a fully Lagrangian method for
fluid dynamics. It has been demonstrated to be robust and flexible
({\it e.g.} Review by \cite{M92}). SPH has the advantage of following
collapse of structure with constant mass resolution. Eulerian codes
have high gas resolution in voids, but structures arising from small
scale perturbations in the early universe are limited to the mass
resolution, which is similar in both methods. Eulerian codes without
adaptive refinement smooth structure in collapsing regions.

We use a predictor-corrector time stepping scheme allowing us to vary
the global timestep. The code runs very quickly at first, but slows when the
particle-particle section of the gravity solver becomes dominant. 

The code includes radiative cooling and photoionization heating with
equilibrium abundances. The species we consider are H, H$^{+}$, He,
He$^{+}$, He$^{++}$ and e$^{-}$.

We make use of a highly stable, 2nd order implicit scheme for the
energy equation to avoid excessively small timesteps associated with
the heating/cooling timescales. We find an iterative solution for the
energy, $E$,
\begin{eqnarray}
\frac{E(t+\Delta t)-E(t)}{\Delta t}=PdV+(Heat-Cool)\left(T(\frac{E(t)+E(t+\Delta t)}{2}),\rho\right). \nonumber 
\end{eqnarray}
The predicted midpoint values for $\rho$ and the $PdV$ work are used.

\subsection{{The P$^{3}$MG Gravity Solver}}

Iterative Full Multigrid techniques provide an excellent alternative
to Fourier based PM Methods. Though comparable in speed for a periodic
computation, the Multigrid method requires less work for free boundary
calculations, by roughly a factor of two.  In addition, we take
advantage of previous time-step information to increase speed even
further by reducing the number of iterations required for convergence.
We employ a multipole expansion of the particle potentials to provide
boundary conditions for the free boundary calculation.  We have added
Particle-Particle interactions to dramatically augment the resolution
of our gravity solver. The resulting P$^{3}$MG code achieves $\sim$1\%
force errors.

For these simulations we have multiple particle masses. We make use of
nested grids to optimise our computational costs. We use a $128^3$
grid with P-P around the high resolution core region with an
additional $64^3$ grid with no P-P forces for the lowest resolution
particles whose function is only to provide long range tidal forces.
This provides for a maximum resolution of 1 kpc. We set this
as the resolution limit for our gas and gravity.

\subsection{Enhancements to Standard SPH}

SPH relies on weighted sums over near neighbours to
estimate all fluid quantities including the pressure forces that move
the particles. The method is adaptive, with the smoothing length for
the weighted sum being a direct measure of the local resolution.

Appropriately varying smoothing lengths, $h$, are essential to any SPH
implementation. Methods based on numerically estimated quantities such
as divergence or density are susceptible to noise whereas what we
require is a good sample of neighbouring particles within $2\,h$.  To
this end we count neighbours in three equi-volume shells around each
particle, $(0-1.58)\,h_{old}$, $(1.58-2)\,h_{old}$ and
$(2-2.29)\,h_{old}$.  We update $h$ by interpolating over these bins
to have 36 neighbours within $2\,h_{new}$. We perform the counting
during the density summation (to avoid extra computation) for use
during the next timestep.

The SPH method involves only pairwise interactions, giving exact
momentum and energy conservation. It also relies upon explicit
artificial viscosity to model shocks correctly. The artificial
viscosity for particle $i$, contributed by a neighbour $j$, is a
function of $(\vec v_{i}-\vec v_{j})\cdot(\vec r_{i}-\vec r_{j})$.
This form cannot differentiate between truly converging flows and pure
shear flows.  Numerical viscosity is thus indiscriminately applied
unless a switch is used, such as $|div|/(|div|+|curl|+\epsilon)$ ({\it
e.g.} \cite{S96}). Rather than finding the exact divergence and
vorticity to flag particles in strongly convergent (shocking) regions,
we use a similar form that is easier to calculate,
\begin{eqnarray}
Switch_{i}=\frac{\sum_{j} W_{ij} m_{j} 
(\vec v_{i}-\vec v_{j})\cdot(\vec r_{i}-\vec r_{j})}
{\sum_{j} W_{ij} m_{j} \mid(\vec v_{i}-\vec v_{j})
\cdot(\vec r_{i}-\vec r_{j})\mid} \, . \nonumber 
\end{eqnarray}

A linked-list of particles allocated to a grid of cells is an
extremely fast method for locating particle neighbours.  Tree
structure is an alternative method, though roughly six times slower for
uniform densities and very memory intensive.  However, with high
density contrasts, many particles may accumulate in a few link list
cells, rendering the method locally ${\cal O}(n^2)$ and thus very
slow. We adaptively refine our linked list grid to avoid this problem
- to a maximum level of four binary sub-divisions. The storage
requirements are naturally much larger but the resulting method slows
by only a factor of three in extremely clustered arrangements.

\section{The Simulations}\label{IC}

\begin{figure} 
\plottwo{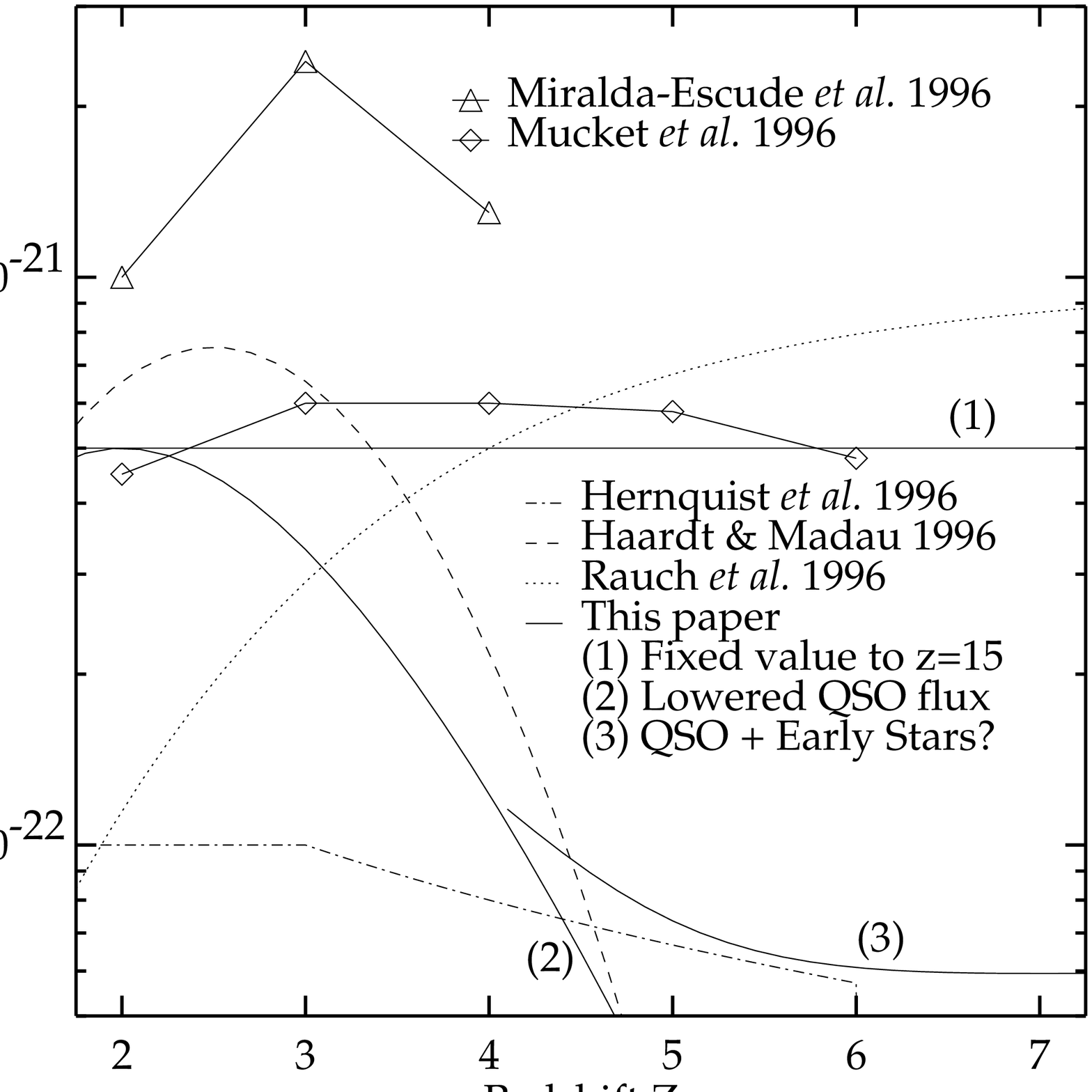}{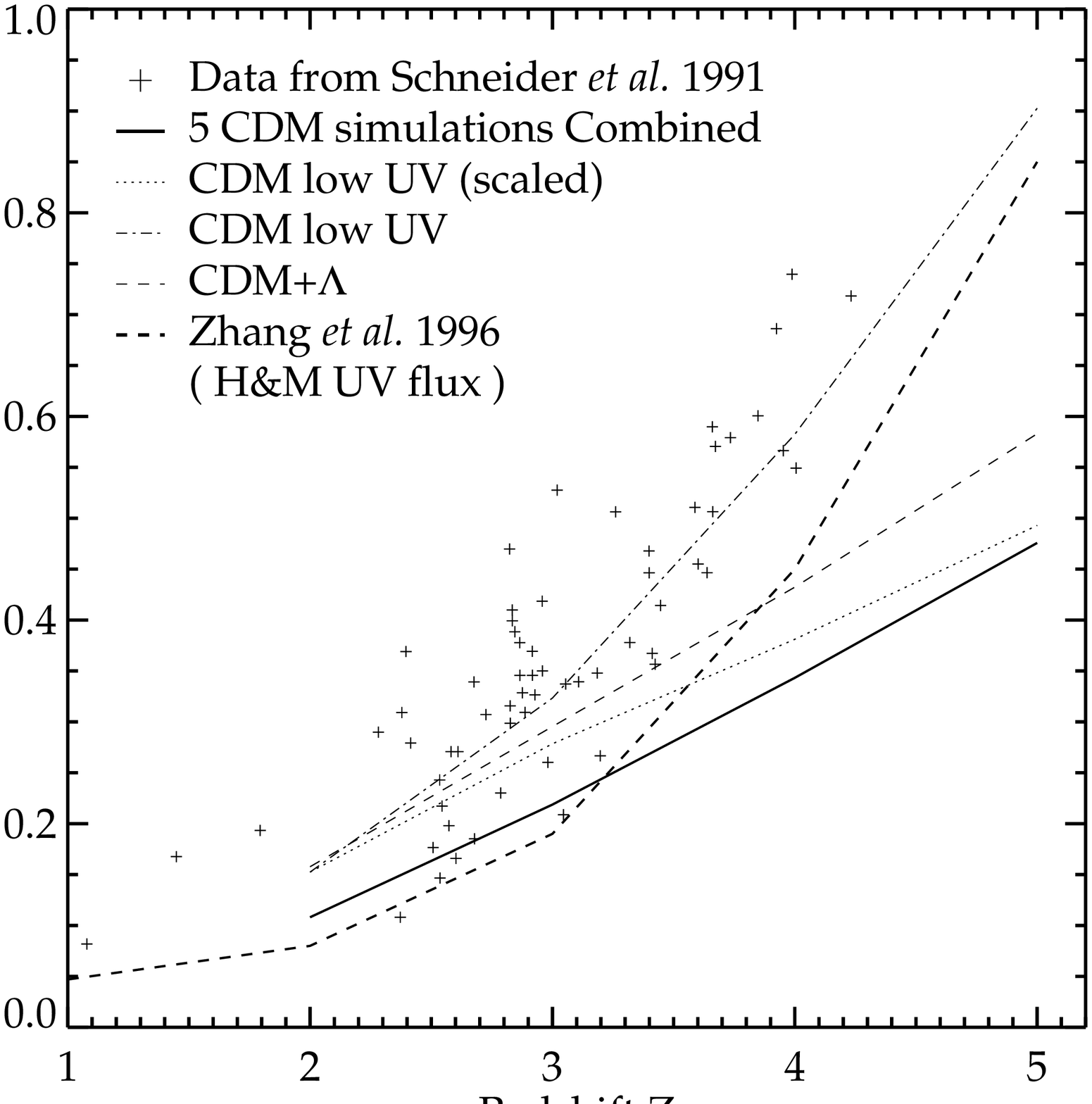}
\caption{\scriptsize UV Flux history and Continuum Depression,
$D_A$. Curve (2) in the left hand panel was designed to reproduce the
$D_A$ vs. redshift data shown in the right panel. A curve such as (3),
including an estimate of the UV flux from early stars, is also
consistent with the $D_A$ data. The flux model of Haardt \& Madau (1996) is
too high for standard CDM models, as demonstrated by the dashed curve
in the right hand panel.}\label{UVDA}
\end{figure}

The photoionization levels and heating are determined by the
background UV flux, in particular the flux at the ionization edge for
hydrogen, $J_{912}$. Observations of the proximity effect hint at
little evolution in $J_{912}$ over the range $z=2$ to $4$ ({\it e.g.}
\cite{B94}). As shown in the left panel of Figure~\ref{UVDA}, we apply
both a fixed flux (1) and a strongly evolving flux similar to that of
\cite{HM96} lowered to fit the continuum depression data (2), given by the
functional form, $5\times 10^{-22}\,((1+z)/3)^9\,\exp(-3\,(z-2))$. These
choices should roughly bracket the true mean UV flux. 

The small scale nature of the clouds demands high resolution numerical
work, especially as the collisionless nature of the dark matter plus
radiative cooling prevent the classical Jean's length,
$k_J^{-1}\approx 0.049 (4/(1+z)\,(T/10^{4}\,K))^{1/2}\,\Omega_{nr}^{-1/2}$
h$^{-1}$ Mpc from setting a minimum scale. $\Omega_{nr}$ refers to
density parameter in non-relativistic matter.

We simulate regions of comoving diameter $5.0 {\rm\,Mpc}$ ($2.5 {\rm\,
h^{-1}Mpc}$ for the CDM model, $3.5 {\rm\, h^{-1}Mpc}$ for the
CDM+$\Lambda$ model) with $50^3/2$ gas and $50^3/2$ dark matter
particles at our highest resolution. We enclose this in a medium
resolution buffer of gas and dark matter particles with 8 times the
mass, out to $8.0{\rm\,Mpc}$ and then massive tide particles with 64
times the mass, out to $12.8 {\rm\,Mpc}$. We apply a linearly evolved
external shear to the entire volume, derived from the initial
conditions. Thus we make considerable effort to model the long wave,
tidal environment well.

We use a sophisticated constrained field code to set up the initial
displacements from unperturbed lattice positions, using the Zeldovich
approximation. It allows for arbitrary types and numbers of
constraints. Our initial conditions finely sample an enormous range of
$k$-space, down to $k=0.01\,h Mpc^{-1}$ for these simulations (see
Figure~1 of BW). Periodic simulations must truncate the large scale
power beyond the scale of the fundamental mode of the box and the
power spectrum is still very flat at $k\sim 1$ h\,Mpc$^{-1}$, the
scale of Lyman-$\alpha$ simulations.

Most simulations use random regions of the Universe. We prefer to have
control parameters which govern some of the major large scale
characteristics of the simulation volume. We constrain the density (via
$\nu$), bulk flow and tidal fields smoothed over three Gaussian filtered
scales $R_f$, where $\nu = \delta_{L,pk}/\sigma_{R_f}$ the overdensity
of the region relative to the {\it rms} fluctuation level
$\sigma_{R_f}$. More details and applications of our peak constraint
approach are described in BW. For these simulations, we fix $\nu$ for
$R_f=0.5$ Mpc and use mean field expectations to set the other
parameters, and the values at two other scales, 1.0 and 1.5
${\rm\,Mpc}$. The $\nu$'s used are shown in Table~\ref{table}. Rare
events which form large ``bright'' galaxies by $z\sim 3$ in the patch
require higher $\nu$. $\nu$ has a Gaussian probability distribution,
allowing us to appropriately weight and combine simulations of
different $\nu$ to create a sample broadly representative of the
universe. In creating samples of Lyman-$\alpha$ statistics, the solid
angle on the sky subtended and length of spectra generated by the
region add further weighting.

\begin{table}
\caption{Initial Conditions used in this work.}
\begin{center}\scriptsize
\begin{tabular}{llcllll} 
Cosmology & {\it Description} & $\nu$ & $e_{v},\,p_{v}$\tablenotemark{a} & Weight\tablenotemark{b} & Angle\tablenotemark{c} & $\Delta\,\lambda$\tablenotemark{c} \\
\tableline 
CDM & Void & -1.4 & 0.179, 0.014 & 0.069 & 1.47 & 1.46 \\ 
 & Weak Void & -0.7 & 0.337, 0.034 & 0.242 & 1.26 & 1.36\\
 & Flat &  0.0 & 0, 0 & 0.379 & 1.0 & 1.0 \\
 & Weak Peak &  0.7 & 0.337, 0.034 & 0.242 & 0.78 & 0.77 \\
 & Peak & 1.4 & 0.179, 0.014 & 0.069 & 0.57 & 0.62 \\ 
CDM+$\Lambda$ & Flat & 0.0 & 0, 0 \\
\end{tabular}
\end{center}
\tablenotetext{a}{Tidal (shear) field parameters}
\tablenotetext{b}{Weighting due to $\nu$ only}
\tablenotetext{c}{Solid angle and average length of spectra, $\Delta\,\lambda$, measured at $z=3$ relative to Flat region}
\label{table}
\end{table}

We describe results for two cosmological models: standard CDM,
normalised to $\sigma_{8}=0.67$ with ${\rm h}=0.5$, $\Omega_{B}=0.05$;
COBE-normalised CDM+$\Lambda$, with ${\rm h}=0.7$,
$\Omega_{B}=0.0255$ and $\Omega_{nr}=0.335$ (shape parameter
$\Gamma=0.225$ and tilt $n_s=0.94$, which gives a theory in agreement
with the shape of the observed large scale power spectrum, see
Figure~1 of BW).  We required that $\sigma_{0.5{\rm\,Mpc}} = 1.05$ at
$z=3$ and the number of baryons in the simulation volume to be the same in
both models.

\section{Results}\label{results}

To analyse our simulations we produce artificial spectra, with signal
to noise of 100 and 5~km\,s$^{-1}$ pixels. We sample a regular spatial
grid of lines of sight running along the three axial directions
through our simulations. A typical line of sight is $\sim$500
km\,$s^{-1}$ long and a single simulation sample $\sim10^{6}$
km\,$s^{-1}$ at z=3.  We fit Voigt profiles using an automated profile
fitting program, designed to emulate the methods employed by
observers. The program identifies each group of lines as a region in
the spectrum where the flux drops below 98\% of the
continuum. $\chi^2$ minimization is used to fit first one line, then
two lines and so on. The number of lines that produces the lowest {\it
reduced}-$\chi^2=\chi^2/d$ is used. $d$ is the number of degrees of
freedom remaining for the fit, estimated as 1 per 2 pixels in the line
group region minus 3 for each line used in the fit. In practice a line
group rarely needs more than 6 lines for a good fit.

\begin{figure}
\plotone{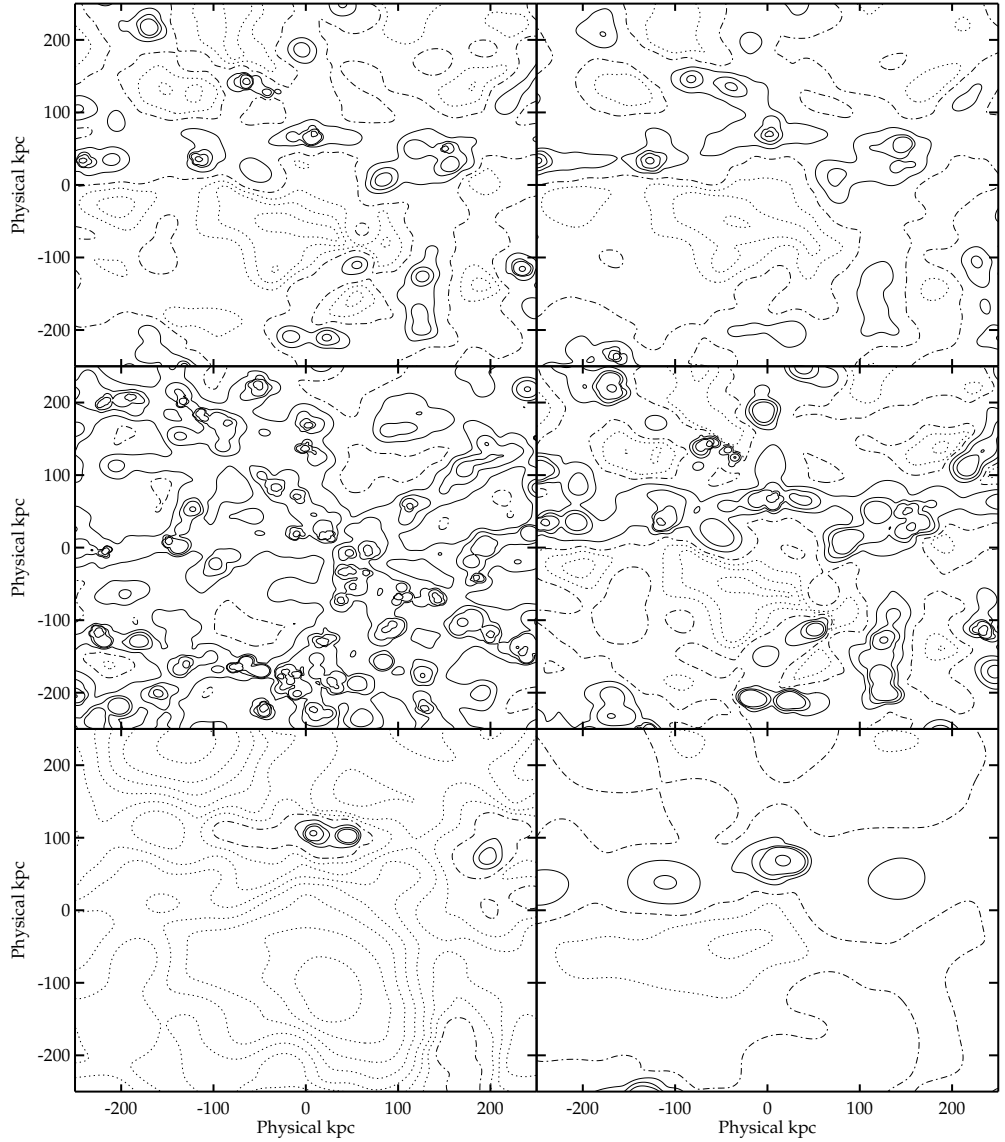}
\caption{\scriptsize HI Column depth through a cube 2 comoving Mpc
thick at z=3.  The contour levels, in log$_{10}$(cm$^{-2}$), are
11,11.25,11.5,11.75,12,12.25,12.5,12.75,13 ({\it dotted}) ,13.36 (mean
density, {\it dot dashed}) and 14,14.5,15,17 ({\it solid}). The left
panels are CDM simulations, with $\nu=0$, $+1.4$ and $-1.4$, top to
bottom. Note the incredible enhancement of the filamentary structure
in the overdense regions. The right hand panels are $\nu=0$ runs
directly comparable with the CDM run in the top left panel. In the top
right panel we show a CDM+$\Lambda$ run, which has more large scale
power and thus more enhanced filaments. The run in the right centre
panel had a lower UV flux at early times, resulting in a more clumpy
distribution and greatly enhanced HI column depths. The lower right
panel shows that poor resolution (1/2 the spatial resolution of the
upper left panel) will remove much of the structure.}
\label{crosssection}
\end{figure}

Figure~\ref{crosssection} demonstrates visually the difference between
several runs, as shown in the column depth of neutral hydrogen through
the central 2 Mpc of the simulations.

\begin{figure}
\plotone{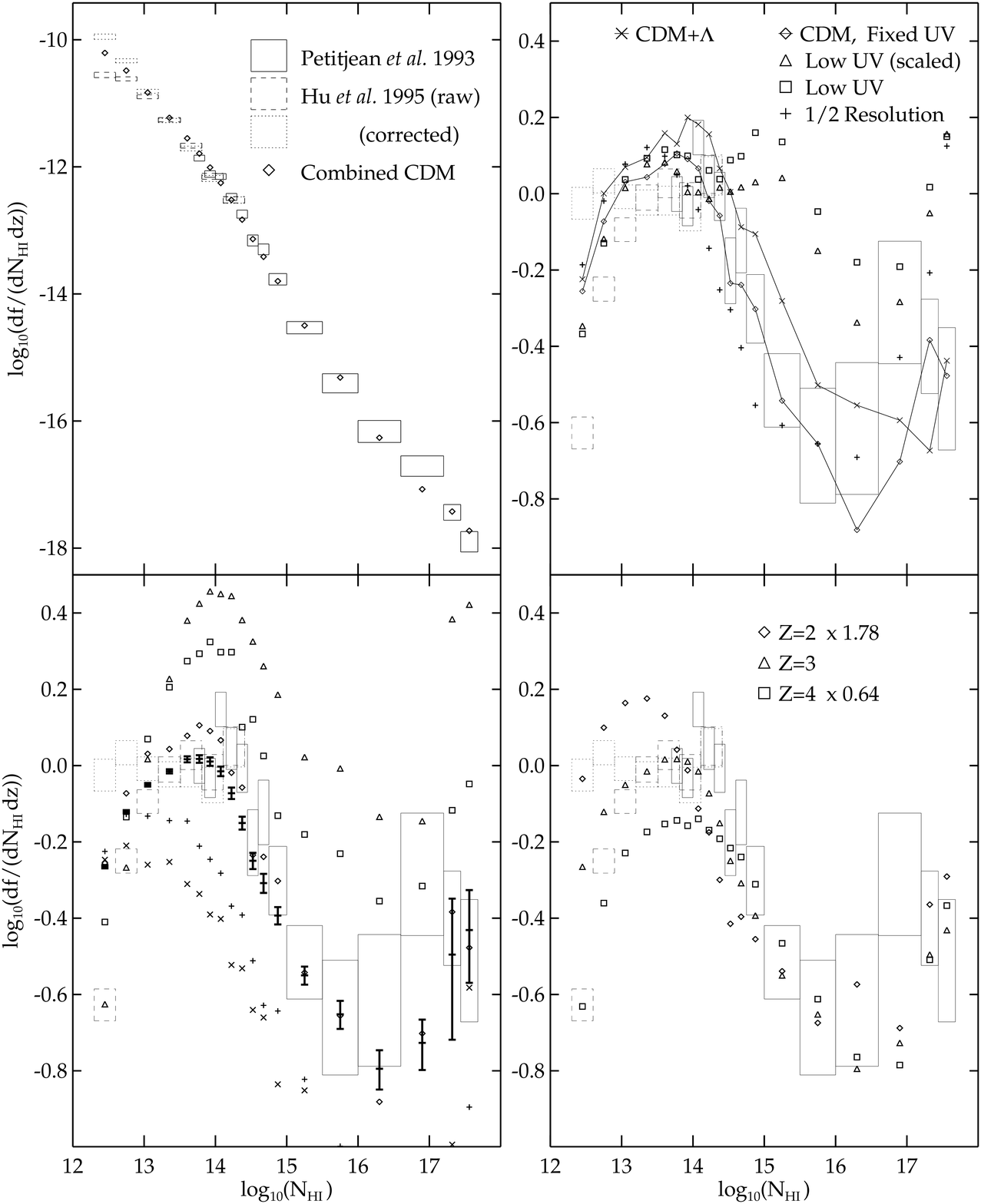}
\caption{\scriptsize Frequency distribution of lines by $N_{HI}$.  In
all frames the boxes represent binned data, box width equalling bin
width and the upper and lower edges denoting 1 sigma Poisson errors.
At the top left the full range in frequency is plotted and in the
remaining graphs, the power law $-1.46\,log_{10}(N_{HI})+8.25$ has
been subtracted for clarity. On the left, the data from 5 CDM
simulations are plotted.  The symbols in the lower plot represent
(from the top down) the $\nu=1.4$, $0.7$, $0$, $-0.7$ and $-1.4$ runs
respectively and the combined data are shown with 1 sigma Poisson
error bars. At the top right, the effects of various choices for the
UV flux, numerical resolution and cosmology are compared. The low UV
flux at early times results in more clumpiness in the IGM,
overproducing lines at intermediate column depths. At the bottom
right, the data resulting from combining the CDM simulations are shown
at 3 redshifts, divided by $((1+z)/4)^2$. The evolution inferred  from
the observations is closer to $(1+z)^3$, suggesting a UV flux
decreasing with redshift may be needed. } \label{fNHI}
\end{figure}

The left panels show the dramatic effect of varying $\nu$ and hence the
shear field. The filamentary structure is greatly enhanced in even
slightly overdense regions. This has important repercussions in the
types and numbers of Ly$\alpha$ absorption lines produced, as shown in
the bottom left panel of Figure~\ref{fNHI}. Simulating overdense
regions or voids like these is not possible in a periodic box of a
similar size, because the density averaged over the box must remain
equal to the universal mean value. Even for significantly larger
boxes this is a problem.

Comparing the CDM and CDM+$\Lambda$ runs (top left and right panels in
Figure~\ref{crosssection} respectively), it can be seen that the
flatter spectrum in the CDM+$\Lambda$ case makes the filaments more
prominent and the dwarf galaxies less so, but without it having a
major impact on the $N_{HI}$ frequency curve. The difference in
normalisation is due to the width in redshift of the simulation volume
in each cosmology, $dz/dl_{com}=H_{0}/c (\Omega_{nr} (1+z)^3 +
\Omega_{\Lambda})^{1/2}$.  The high resolution size is $\Delta
l_{com}$=5 Mpc, comoving, in all simulations. Dividing by the ratio of
$\Delta z=dz/dl_{com} \Delta l_{com}$ at z=3 moves the CDM+$\Lambda$
curve down by 0.09 on the top right panel of Figure~\ref{fNHI}.

The left hand panels of Figure~\ref{fNHI} demonstrate that an
appropriate fixed choice of $J_{912}=5\times\,10^{-22}$ can reproduce
the observed statistics very well. We have statistically combined 5
CDM simulations to effectively produce a good sample of the universe
that includes rarer regions. The advantage of this over a single large
simulation is great resolution and importance sampling. The
contribution of different regions is apparent. The voids, $\nu<0$,
cause a lowering of the curve, especially at low column depths. 

The results of a low resolution run (half that of the standard ones)
are shown represented by crosses in the top right panel of
Figure~\ref{fNHI}. Though it underpredicts the low column lines, the
high column lines are overpredicted relative to our higher resolution
runs: limited resolution mimicks heating by ``puffing'' up the cores of cold
lumps out to greater sizes and thus gives higher cross sections.

In the top right panel of Figure~\ref{fNHI} we demonstrate that
attempts to renormalise simulations with inappropriate values of the
ionizing flux will fail to give the correct frequency of $N_{HI}$.
The curves of interest are the CDM line with a fixed UV history and
the scaled low UV points (triangles). These simulations are otherwise
identical and have been processed with a fixed value of
$J_{912}=5\times 10^{-22}$. What {\it does} scale fairly well is the
continuum depression, $D_A$, shown in the right panel of
Figure~\ref{UVDA}. $D_A$ is defined at the mean value of
$1-\exp(-\tau)$ for a spectral region.  This is because the dominant
contributors to the optical depth are lines with $N_{HI} \sim
10^{14}$, resulting from barely overdense material for which the UV
flux history is less important.

The past history of the UV flux is important for the formation of
dense gaseous clumps, as illustrated in the top left and centre right
panels of Figure~\ref{crosssection}; the same UV flux was applied for
the post-processing analysis, but the simulation shown in the centre
right panel experienced a low UV flux during its evolution (curve (2)
in Figure~\ref{UVDA}). Applying a different flux in post-processing
the calculation can of course have no effect on the density
distribution and thus cannot compensate for a different history.

Subtracting a power law, derived from the \cite{Hu95} low end slope,
from the $f(N_{HI})$ curve of Figure~\ref{fNHI} reveals interesting
features in the data. If the bumps at $N_{HI}=10^{14}$ and
$N_{HI}=10^{17}$ (where self shielding of the UV flux starts to
become important) are real, they may allow simulations to finely
discriminate between parameter choices. The data appear to follow a
power law slope of --1.9 in the range
$N_{HI}=10^{14}-10^{16}\,$cm$^{-2}$, which our runs with fixed flux
reproduce and those with low UV flux at early times do not. This is
also apparent in the results of \cite{R96} who apply a UV flux that
slowly increases with redshift. Simulations that calculate the UV flux
history (\cite{ME96}, \cite{M96}) perform fairly well, suggesting that
a more detailed self-consistent treatment could produce a strong
result. The excess of intermediate column lines seen for the low UV
simulations may be due to unrealistically low energy input at early
times; specifically there should be input from early stars. We are
testing this hypothesis using a UV flux history similar to curve (3)
in Figure~\ref{UVDA}. Star formation and supernova feedback would also
inhibit the formation of the extremely dense, cool clumps associated
with this excess. We are now incorporating these processes using a
general chemical evolution code.

{\it We have illustrated that the Lyman alpha forest is a sensitive
probe of the conditions that operated at $z \sim 5$ to $2$. It is
clear that we must produce a comprehensive sample if we are to deliver
reliable predictions to the observers of the clouds. Finding forest
observables that allow us to separate the influence of primordial
power spectrum shape and amplitude, cosmological parameters such as
$\Omega_\Lambda$, $\Omega_{cdm}$, $\Omega_{hdm}$, $H_0$ and
$\Omega_B$, the ionization history and local energy injection from
galaxies (feedback) will be a challenge to the growing group of cloud
simulators, in close collaboration with the observers of the
forest. Nonetheless it is clear that the basic model (whether CDM or
CDM+$\Lambda$) with $\sigma_{0.5 {\rm\,Mpc}} \sim 1$ at $z\sim 3$
provides a rather good fit to the data.}

\acknowledgments

Support from the Canadian Institute for Advanced Research and NSERC is
gratefully acknowledged. We would like to thank John Lattanzio and Joe
Monaghan, who were involved in the early phases of this project, using
an SPH-Multigrid code of earlier vintage.


\begin{thebibliography}{}

\bibitem[Bechtold (1994)]{B94}
\reference Bechtold, J. 1994, in {\it QSO Absorption Lines}, ESO Astrophysics Symposia, G. Meylan (ed), Springer-Verlag, p.299

\bibitem[Bond \& Wadsley (1996)]{BW} \reference Bond, J.R. \&
Wadsley, J.W. 1996, to be published in {\it Computational
Astrophysics}, 12th Kingston Meeting on Theoretical Astrophysics, {\it
in preparation}, (BW)

\bibitem[Dav\'{e} {\it et al.} (1996)]{dave96}
\reference Dav\'{e}, R., Hernquist, L., Weinberg, D. \& Katz, N. 1996, Accepted by \apj

\bibitem[Haardt \& Madau (1996)]{HM96}
\reference Haardt, F. \& Madau, P. 1996, \apj, 461, 20

\bibitem[Hernquist {\it et al.} (1996)]{H96}
\reference Hernquist, L., Katz, N., Weinberg, D. \& Miralda-Escud\'{e}, J. 1996, \apj, 457, L51

\bibitem[Hu {\it et al.} (1995)]{Hu95}
\reference Hu, E.M., Kim, T.S., Cowie, L.L., Songaila A. \& Rauch, M. 1995, \apj, 466, 46

\bibitem[Miralda-Escud\'{e} {\it et al.} (1996)]{ME96}
\reference Miralda-Escud\'{e}, J., Cen, R., Ostriker, J.P. \& Rauch, M. 1996, \apj, 471, 582

\bibitem[Monaghan (1992)]{M92}
\reference Monaghan, J.J. 1992, \araa, 30, 543

\bibitem[M\"{u}cket {\it et al.} (1996)]{M96}
\reference M\"{u}cket, J.P., Petitjean, P., Kates, R.E. \& Riediger, R., 1996 \aa, 308, 17

\bibitem[Petitjean {\it et al.} (1993)]{PJ93}
\reference Petitjean, P., Webb, J.K., Rauch, M., Carswell, R.F. \& Lanzetta, K. 1993, \mnras, 262, 499

\bibitem[Rauch {\it et al.} (1996)]{R96}
\reference Rauch, M., Haehnelt, M.G. \& Steinmetz, M. 1996, Submitted to \apj

\bibitem[Steinmetz (1996)]{S96}
\reference Steinmetz, M. 1996, to be published in {\it The Early Universe with the VLT}, ESO Workshop

\bibitem[Zhang {\it et al.} (1996)]{Z96}
\reference Zhang, Y., Anninos, P., Norman, M.L. \& Meiksin, A. 1996, submitted to \apj
\end{thebibliography}
\end{document}